\documentstyle[12pt]{article}

\newcommand{\spone}{0.9}

\newcommand{\spthree}{2.4}

\newcommand{\singlespace}{\edef\baselinestretch{\spone}\Large\normalsize}

\newcommand{\threespace}{\edef\baselinestretch{\spthree}\Large\normalsize}
\threespace
\begin{document}
\singlespace

\begin{center}
{\bf A Unified Construction of Variational Methods for the Nonlinear \\
Schr\"odinger Equation}  \\
\vspace{8pt}

{ \bf Yeong E. Kim and Alexander L. Zubarev} \\
\vspace{5pt}

Department of Physics. Purdue University \\
West Lafayette, IN   47907 \\
\end{center}
\vspace{8pt}

Based on an approach introduced by Gerjuoy, Rau, and Spruch, we construct
variational principles in a systematic way for the nonlinear Schr\"odinger
equation and obtain new variational principles for the case of 
Ginzburg-Pitaevskii-Gross equation (GPG) which is believed to describe
accurately the Bose-Einstein condensation at zero temperature.  As an
application of these variational methods, a variational iteration
method is proposed for calculating eigenvalue (chemical potential) and wave function
for the GPG equation
\vspace{25pt}

\noindent
PACS number(s): 03.75.Fi, 05.30.Jp, 67.90.+z
\vspace{8pt}

\noindent
{\bf 1.  Introduction}
\vspace{8pt}

The concept of the Bose-Einstein condensation (BEC) [1] has been
known for 73 years, and has been used to describe all physical scales,
including liquid $^4He$, excitons in semiconductors, pions and kaons in dense
nuclear matter (neutron stars, supernovae), and elementary particles [2].
It is only a few years ago that the BEC phenomenon
was observed directly in dilute vapors of alkali atoms, such as rubidium [3],
lithium [4], and sodium [5], confined in magnetic traps and cooled down
to nanokelvin temperatures.  
\vspace{8pt}

These remarkable experimental observations [3-5] have stimulated much
activities in experiments [6-11] and in the theoretical investigation 
of inhomogenous Bose gases.  It is impractical to quote all of numerous
theoretical papers published during the past few years.  We quote here only
a fraction [12-28] of the theoretical papers, which are relevant to this
paper and are 
based mostly on the Ginsburg-Pitaevskii-Gross (GPG) equation (also known as
Pitaevskii-Gross (PG) equation) [29-32], which
is believed to describe accurately the experimental results well below
the onset of BEC.  The GPG equation is a nonlinear Schr\"odinger equation
(NLSE), which is formally identical to a Ginzburg-Landau type of field
equations (the Ginzburg-Landau theory is also known as $\psi$ theory [36]). 
The NLSE is one of the nonlinear differential equations,
the study of which led to fundamental advances in nonlinear dynamics [37].

Since the vapors used in the experiment are dilute, the average distance
between the atoms is substantially larger than the range of the
interaction.  Hence, the Bose gas is expected to be dominated by
two-body collisions, which can be well described by the s-wave scattering
length.  Thus, the Ginsburg-Pitaevskii-Gross theory developed for weakly
interacting bosons is ideally suited to describe BEC observed in the
experiments. 
\vspace{8pt}

The numerical methods based on variational 
principles (VP) are the most effective approximation method and have been
used to solve the GPG equation
using the numerical minimization of the
energy functional [14, 16, 21, 25, 28].  However, it is expected
that the wave function 
determined by the variational method is much less accurate than 
the stationary value of the functional itself (e.g. energy),
because the later value is only indirectly related to the wave function
and depends weakly on the trial wave function near the stationary
point.  We note in contrast that, for the case of the linear
Schr\"odinger equation, one can obtain the criteria of accuracy
not only for stationary quantities but also for the wave function
using Schwinger variational principles [35].
\vspace{8pt}

In this paper, we present a systematic construction of VP
which is free of the above shortcomings, based on the approach
proposed by Gerjuoy, Rau, and Spruch (GRS) [36].  We introduce a functional
whose optimal value is the wave function at given values of its
arguments. In section 2, we give a brief review of the GPG (or PG)
equation as it relates to BEC.  In section 3, we give an outline of GRS
approach for variational principles.  Section 4 describes our
derivation of a functional,
stationary value of which is the chemical potential, and section 5
describes the VP for the wave function.  In
section 6, we describe our formulation of iteration method based on
our VP for the wave function.  Summary and
conclusions are given in section 7.   In the Appendix we describe
an example of the application of our variational iteration method to
a nonlinear integral equation.
\vspace{8pt}

\noindent
{\bf 2.  Nonlinear Schr\"odinger Equation} 
\vspace{5pt}

For a dilute gas of bosons at zero temperature mean field theory
yields a nonlinear Schr\"odinger equation (NLSE) for the condensate
wave function [32].  The condensate wave function (or order parameter)
may be described by a self-consistent mean field NLSE known as the
Ginsburg-Pitaevskii-Gross (GPG) equation (also known as Gross-Pitaevskii
(GP) equation [29-31].  For condensed neutral atoms in an external
trap, the GPG equation takes a form 
$$ \frac{i \hbar \partial\psi(\vec{r},t)}{\partial t} =~ \frac{\hbar^2}{2m}~
\nabla^2 \psi(\vec{r},t) + V_{trap}(\vec{r}) \psi(\vec{r},t) + V_H(\vec{r})
\psi(\vec{r}, t)
\eqno{(1)}
$$

\noindent 
where $\psi(\vec{r},t)$ is the BEC wavefunction,$m$ is the atom mass, 
and $V_{trap}(\vec{r})$ is a trap potential.  $V_H(\vec{r})$ is the mean
Hartree interaction energy for the condensate
$$ V_H(\vec{r}) = g_0 N_0 |\psi(\vec{r},t)|^2 ,
\eqno{(2)}
$$

\noindent
where $N_0$ is the total m number of the BEC atoms, and
$$ g_0 = \frac{4 \pi \hbar^2a}{m}
\eqno{(3)}
$$

\noindent
with the scattering length $a$.  To obtain a stationary solution, we insert
$\psi(\vec{r}, t) = exp (-i \frac{\mu_0}{\hbar}t) \psi(\vec{r})$ (where $\mu_0$ is the
chemical potential) into Eq. (1) to find the following equation for
$\psi(\vec{r})$
$$ - \frac{\hbar^2}{2m} \nabla^2\psi(\vec{r}) + V_{trap}(\vec{r})
\psi(\vec{r}) + g_0 N_0 |\psi(\vec{r})|^2 \psi(\vec{r}) = \mu_0 \psi(\vec{r}) ,
\eqno{(4)}
$$

$$ \int d \vec{r}|\psi (\vec{r})|^2 = 1  .
\eqno{(5)}
$$

\noindent
For an anisotropic harmonic oscillator trap, $V_{trap}(\vec{r})$ is given
by
$$V_{trap}(\vec{r}) = m(\omega_x^2 x^2 + \omega_y^2 y^2 + \omega_z^2 z^2)/2.
\eqno{(6)}
$$

\noindent
In this paper, we develop the variational principles, for solving Eq.
(4) based on the approach proposed by Gerjuoy, Rau, and Spruch [36].  The GRS
variational principles for the time-dependent Eq. (1) will be given elsewhere.
\vspace{8pt}

\noindent
{\bf 3. Variational Principles of Gerjuoy, Rau, and Spruch}.
\vspace{5pt}

In this section, we give a brief outline of the unified formulation of the
construction of variational principles, developed by Gerjuoy, Rau, and
Spruch GRS [36].   
\vspace{5pt}

We start with the following set of equations 
$$ B_i(\phi) = 0,~~ i = 1, 2,...,
\eqno{(7)}
$$

\noindent
where $\phi$ designates a set of function are precisely defined by
relations (7).
\vspace{5pt}

Now, let $(\phi_t, \phi_t^\ast)$ be a trial estimate of exact but unknown
$(\phi, \phi^\ast)$, and let $(L_t^i, \tilde{L}_t^i),$ be a trial
estimate of a $({\it L}^i, \tilde{L}^i)$ to be specified.  Then, the
functional [F] = $F_v(\phi_t, \phi_t^\ast, \it {L}, {\it \tilde{L}})$
is a variational functional for the desired $F(\phi, \phi^\ast)$ which we seek, if 
$$ F_v (\phi, \phi^\ast, {\it L}_t, {\it \tilde{L}}_t) = F(\phi, \phi^\ast),
\eqno{(8)}
$$

\noindent
and
$$ \delta[F] = F_v(\phi_t, \phi_t^\ast, {\it L}_t, {\it \tilde{L}}_t) -
F(\phi, \phi^\ast) = O(\delta \phi^2, \delta {\it L}^2, \delta \phi \delta
{\it L}), 
\eqno{(9)}
$$

\noindent
where
$$ {\it L}^i = {\it L}_t^i + \delta {\it L}^i,~ {\it \tilde{L}}^i = 
{\it \tilde{L}}^i + \delta {\it \tilde{L}}^i,~ \phi = \phi_t + \delta \phi,~
\phi^\ast = \phi_t^\ast + \delta \phi^\ast  .
\eqno{(10)}
$$
\vspace{5pt}

\noindent
Now, consider 
$$ [F] = F(\phi_t, \phi_t^\ast) + \sum_i (\langle L_t^i|B_i(\phi_t)\rangle + \langle B_i(\phi_t) \tilde{L}_t^i \rangle)  .
\eqno{(11)}
$$

\noindent
Using Eqs. (9 - 11) we can determine the necessary relations specifying
$(L^i, \tilde{L}^i)$, whose estimates $({\it L}^i, {\it \tilde{L}}^i)$ appear in Eq. (9).  The essential point is that there are no
restrictions on the nature of $\phi, F(\phi)$, and $B(\phi)$.  Therefore
the GRS method is a general method for construction of VP for linear,
nonlinear, differential, and integral equations.  
\vspace{8pt}

\noindent
{\bf 4.  Variational Principle for the Chemical Potential}
\vspace{5pt}

In this section we derive a functional whose stationary value is the chemical
potential $\mu_0$.  We start with Eq. (4)
$$(T + V_{trap} + g_0 N_0 |\psi|^2)\psi = \mu_0 \psi,
\eqno{(4a})
$$

\noindent
where T is the kinetic energy operator 
$$ T= - \frac{\hbar^2}{2m} \nabla^2 .
\eqno{(12)}
$$

\noindent
Now, let us consider a functinal [$\mu_0$]
$$ \begin{array}{rcl}
 [\mu_0] &=& \mu_t + \lambda^t [(\langle \psi_t|\psi_t \rangle - 1) +
\langle L_t|(T + V_{trap} + g_0N_0|\psi_t|^2 - \mu_t)|\psi_t \rangle  \\

 &+& \langle(T + V_{trap} + g_0N_0|\psi_t|^2 - \mu_t)\psi_t|L_t \rangle] ,
\end{array}
\eqno{(13)}
$$

\noindent
where
$$ \lambda^t = \frac{1}{\langle L_t|\psi_t \rangle + \langle \psi_t|L_t \rangle} ,
\eqno{(14)}
$$

\noindent
and $L_t(\vec{r})$ is solution of linear equation
$$ (T + V_{trap} + 2 g_0 N_0|\psi_t|^2)L_t + g_0 N_0 \psi_t^2 L_t^\ast - \mu_t L_t =
- \psi_t,
\eqno{(15)}
$$

\noindent
with boundary conditions
$$ 
 \lim_{r \rightarrow \infty}~L_t(\vec{r}) \rightarrow 0  .
\eqno{(16)}
$$

\noindent
To prove that [$\mu_0$], Eq. (13) is a stationary expression, let us calculate 
$\delta[\mu_0]$, let
$$ \begin{array}{rcl}
  \psi_t &=& \psi + \delta \psi,~ \psi_t^\ast = \psi^\ast + \delta \psi^\ast ,  \\
  \mu_t &=& \mu_0 + \delta \mu,~ L_t = L + \delta L .
\end{array}
\eqno{(17)}
$$

\noindent
We also assume that
$$ \delta \mu = O(\delta \psi),~ \delta L = O(\delta \psi)  .
\eqno{(18)}
$$

\noindent
Substitution of Eq. (17) into Eq. (13) gives
$$\begin{array}{rcl}
\delta[\mu_0] &=& \delta\mu[1 - \lambda(\langle L|\psi \rangle + \langle\psi|L \rangle)] + \\
   \lambda (\langle \delta \psi |\psi \rangle &+& \langle \psi| \delta \psi \rangle +
\langle \delta \psi|(T + V_{trap} + 2 g_0 N_0| \psi|^2 - \mu_o)|L \rangle + \\
   \langle \delta \psi|g_0 N_0 \psi^2|L^\ast \rangle &+& \langle \psi|(T + V_{trap} +
2U_0N_0|\psi|^2 -\mu_0)L|\delta \psi \rangle + \\
   \langle g_0 N_0 \psi^2 L^\ast |\delta \psi \rangle) &+& O(\delta \psi^2) .
\end{array}
\eqno{(19)}
$$

\noindent
Using
$$ \lambda = (\langle L|\psi \rangle + \langle \psi|L \rangle)^{-1} ,
$$       

\noindent
and Eq. (18), we obtain
$$ \delta [\mu_0] = O (\delta \psi^2).
$$

Since homogeneous equation
$$ (T + V_{trap} + 2g_0 N_0|\psi|^2) L + g_0N_0 \psi^2 L^\ast -\mu L = 0 
\eqno{(21)}
$$

\noindent
has a solution
$$ L = ic \psi ,
\eqno{(22)}
$$

\noindent
where c is a real constant, and there is no homogeneous solution in general of
$$ (T + V_{trap} + 2g_0N_0|\psi_t|^2)L_t + g_0N_0 \psi_t^2 L_t^\ast - \mu_t L_t = 0 ,
\eqno{(23)}
$$

\noindent
our assumption given by Eq. (18), $\delta L = O(\delta \psi)$, has to be
justified.  In general, we need to modify Eq. (15). (For the case of the linear
Schr\"odinger equation, see [36]).  However, for ground state of the GPG equation, we
do not need to modify Eq. (15).  In this case, $\psi_t^\ast = \psi_t$, and from
Eq. (15) we write 
$$ \begin{array}{rcl}
 (T &+& V_{trap} + 2g_0N_0|\psi_t|^2)R_eL_t + g_0N_0|\psi_t|^2 R_eL_t - \mu_t R_eL_t =
- \psi_t,  \\
 (T &+& V_{trap} + 2g_0N_0|\psi_t|^2)ImL_t - g_0N_0|\psi_t|^2ImL_t - \mu_tImL_t = 0 ,
\end{array}
\eqno{(24)}
$$

\noindent
where 
$$L_t = Re L_t + i ImL_t .
$$        

\noindent
For this case, we need to solve Eq. (15) for real solution $L_t(L_t^\ast = L_t)$ and
hence the problem will not arise.
\vspace{8pt}

\noindent
{\bf 5.  Variational principle for wave function}
\vspace{5pt}

In this section, we attempt to solve the following problem  how to find variational
estimates of the BEC wave function.  To do this, we introduce a functionals
$[Re \psi](\vec{r})$ and $[Im \psi](\vec{r})$ whose stationar values are
$[Re \psi](\vec{r})$ and $Im \psi(\vec{r})$ respectively
$$ \begin{array}{rcl}
[Re \psi](\vec{r})&=&\frac{1}{2}(\psi_t(\vec{r}) + \psi_t^\ast(\vec{r})) 
-\frac{1}{2}\int L_r^{t^\ast} (\vec{r},\vec{r}^\prime)
[T + V_{trap} + g_0N_0|\psi_t|^2  \\

 &-& \mu_t]\psi_t(\vec{r}^\prime)d \vec{r}^\prime
  - \frac{1}{2}~\int [(T + V_{trap} + g_0N_0|\psi_t|^2 - \mu_t)\psi_t(\vec{r}^\prime)]^\ast
  L_R^t(\vec{r}, \vec{r})d \vec{r}^\prime  ,
\end{array}
\eqno{(25)}
$$

$$\begin{array}{rcl}
[Im \psi] (\vec{r}) &=& \frac{1}{2i} (\psi_t(\vec{r}) - \psi_t^\ast(\vec{r})) - \frac{1}{2i} [ \int L_I^{t^\ast}(\vec{r}, \vec{r}^\prime)[T + V_{trap} + g_0N_0|\psi_t|^2 - \mu_t]
\psi_t(\vec{r}^\prime)d \vec{r}^{\prime} \\

 &+& \frac{1}{2i} \int [(T + V_{trap} + U_0N_0|\psi_t|^2 - \mu_t) \psi_t (\vec{r}^\prime)]^\ast
L_I^t(\vec{r}, \vec{r}^{-\prime}) d \vec{r}^{\prime}]  ,
\end{array}
\eqno{(26)}
$$

\noindent
where
$L_I^t(\vec{r}, \vec{r}^\prime)$, and $L_R^t(\vec{r}, \vec{r}^{\prime})$ are solutions
of linear equations

$$ 
[T + V_{trap} + 2 g_0N_0|\psi_t|^2 - \mu_t]L_R^t(\vec{r}, \vec{r}^\prime) +
g_0N_0 \psi_t^2 L_R^{t^\ast} (\vec{r}, \vec{r}^\prime) = \delta(\vec{r} - \vec{r}^\prime),
\eqno{(27)}
$$

$$[T + V_{trap} + 2 g_0N_0|\psi_t|^2 - \mu_t]L_I^t(\vec{r}, \vec{r}^\prime) - g_0N_0
\psi_t^2 L_I^{t^\ast}(\vec{r}, \vec{r}^\prime) = \delta(\vec{r} - \vec{r}^\prime).
\eqno{(28)}
$$

Varying Eqs. (25) and (26) we obtain
$$ \begin{array}{rcl}
 \delta[Re \psi](\vec{r}) &=& \int~d \vec{r}^\prime 
 \biggr\{  \frac{1}{2} \{ \delta(\vec{r} - \vec{r}^\prime) (\delta \psi(\vec{r}^\prime) + \delta \psi^\ast
(\vec{r}^\prime)) - \\

  &-& \frac{1}{2} [([T + V_{trap} +2g_0N_0|\psi_t|^2 - \mu_t]L_R^t(\vec{r}, \vec{r}^\prime))^\ast \delta \psi(\vec{r}^\prime) \\

  &+& g_0 N_0 \psi^{\ast 2}(\vec{r}^\prime)L_R^t(\vec{r}, \vec{r}^\prime) \delta \psi
(\vec{r}^\prime)  \\

  &+& [T + V_{trap} + 2 g_0N_0|\psi_t|^2 -\mu_t]L_R^t(\vec{r}, \vec{r}^\prime) \delta \psi^\ast
(\vec{r}^\prime) \\

  &+& g_0N_0 \psi^2(\vec{r}^\prime) L_R^{t^\ast} (\vec{r}, \vec{r}^\prime) \delta \psi^\ast
(\vec{r}^\prime)] \biggr \} + O (\delta \psi^2) ,
\end{array}
\eqno{(29)}
$$

\noindent
and
$$\begin{array}{rcl}
\delta[Im \psi](\vec{r}^\prime) &=& \int d \vec{r}^\prime
   \biggr \{ \frac{1}{2i} \{ \delta(\vec{r} - \vec{r}^\prime)(\delta \psi(\vec{r}^\prime)
  - \delta \psi^\ast(\vec{r}^\prime)) \\

 &-& \frac{1}{2i}[([T + V_{trap} + 2g_0N_0|\psi_t|^2 - \mu_t]L_I^t(\vec{r}, \vec{r}^\prime)^\ast \delta \psi(\vec{r}) \\

 &-& g_0N_0 \psi^{\ast 2}(\vec{r}^\prime) L_I^t (\vec{r}, \vec{r}^\prime) \delta \psi(\vec{r}^\prime) \\

 &-& [T + V + 2g_0N_0|\psi|^2 - \mu_t] L_I^t(\vec{r}, \vec{r}^\prime) \delta \psi^\ast(\vec{r}^\prime) \\

 &+& g_0N_0 \psi^2(\vec{r}^\prime) L_I^{t \ast} (\vec{r}, \vec{r}^\prime) \delta \psi^\ast
(\vec{r}^\prime)] \biggr \} + O(\delta \psi^2), 
\end{array}
\eqno{(30)}
$$

\noindent
where we have assumed
$$ \mu_t = \mu_0 + \delta\mu, \delta\mu= O(\delta \psi^2) ,
\eqno{(31)}
$$

\noindent 
and
$$ L_{R,I}^t = L_{R,I} + \delta L_{R,I}, \delta L_{R,I} = O(\delta \psi) .
\eqno{(32)}
$$

\noindent
Substitution of Eqs. (27), (28) into Eqs. (29) and (30) leads to equations
$$ \begin{array}{rcl}
\delta [Re \psi] (\vec{r}) &=& O(\delta \psi^2),  \\

\delta[Im \psi](\vec{r}) &=& O (\delta \psi^2) ,
\end{array}
\eqno{(33)}
$$

\noindent
which prove that Eqs. (25) and (26) are variational functionals.
\vspace{5pt}

Using Eqs. (25) and (26) we can derive a variational principle for $\psi$ itself
$$ \begin{array}{rcl}
 [\psi](\vec{r}) &=& [Re \psi](\vec{r}) + i[Im \psi](\vec{r}) = \psi_t(\vec{r}) + \int~
R_2^{t^\ast}(\vec{r}, \vec{r}^\prime)[T + V_{trap} \\

 &+& g_0N_0|\psi_t|^2 - \mu_t]\psi_t(\vec{r}^\prime) d\vec{r}^\prime \\

 &+& \int~[(T + V_{trap}
+ g_0N_0|\psi_t|^2 - \mu_t)\psi_t(\vec{r}^\prime)]^\ast R_1^t(\vec{r}, \vec{r}^\prime)
d\vec{r}^\prime , 
\end{array}
\eqno{(34)}
$$

\noindent
where
$$ R_1^t(\vec{r}, \vec{r}^\prime) = - \frac{1}{2}(L_R^t(r, r^\prime) - L_I^t(r, r^\prime)),
\eqno{(35)}
$$

$$ R_2^t(r, r^\prime) = - \frac{1}{2}(L_R^t(r, r^\prime)  + L_I^t(r, r^\prime)).
\eqno{(36)}
$$

\noindent
It is convenient to rewrite Eq. (34) as follows
$$ \begin{array}{rcl}
 [\psi](\vec{r}^\prime) &=& -2~\int~\psi_t^\ast(\vec{r}^\prime) g_0N_0|\psi_t(\vec{r}^\prime)|^2
R_1^t(\vec{r}, \vec{r}^\prime) d \vec{r}^\prime \\

 &-& 2~\int~R_2^{t^\ast}(\vec{r}, \vec{r}^\prime)g_0N_0|\psi_t(\vec{r}^\prime)|^2 \psi_t
(\vec{r}) d \vec{r}^\prime  .
\end{array}
\eqno{(37)}
$$

\noindent
Three comments are appropriate here.  First, for the ground state of the GPG equation, we have
$\psi_t^\ast = \psi_t$, and from Eq. (21)
$$ \begin{array}{rcl}
 [\psi](\vec{r}) &=&  \int \psi_t(\vec{r}^\prime) g_0N_0|\psi_t(\vec{r}^\prime)|^2
L_R^t(\vec{r}, \vec{r}^\prime) d \vec{r}^\prime \\

 &+& \int L_R^{t \ast} (r, r^\prime) g_0N_0|\psi_t(\vec{r}^\prime)|^2 \psi_t(\vec{r}^\prime)
d \vec{r}^\prime .
\end{array}
\eqno{(38)}
$$

Secondly, if $L_R(\vec{r}, \vec{r}^\prime) = L_R^\ast(\vec{r}, \vec{r}^\prime)$,
the following equation
$$ [T + V_{trap} + 2g_0N_0|\psi|^2 - \mu_0] L_R + g_0N_0 \psi^2 L_R^\ast = \delta(\vec{r} -
\vec{r}^\prime) 
\eqno{(39)}
$$

\noindent
has a unique solution, since in this case the homogeneous equation has only a trivial solution.

\noindent
Thirdly, if $\psi_t^\ast \not= \psi_t$ (vortex states), Eqs. (27) and  (28)
are meaningless. To see this, we use Eqs. (27), (28), (35), and
(36) to obtain
$$\begin{array}{rcl}
 (T + V_{trap} &+& 2g_0N_0|\psi|^2 - \mu_0)R_2 + g_0N_0 \psi^2R_1^\ast = -\delta(\vec{r} - \vec{r}^\prime),  \\
 (T + V_{trap} &+& 2g_0N_0|\psi|^2 - \mu_0)R_1 + g_0N_0 \psi^2R_2^\ast = 0,
\end{array}
\eqno{(40)}
$$

\noindent
where $\psi$ is exact solution of the GPG equation (4).  Multiplying Eqs.
(40) on the left by $\psi^\ast(\vec{r}^\prime)$ and integrating, we find,
using Eq. (4), that

$$\begin{array}{rcl}
&~& \int~g_0N_0|\psi(\vec{r}^\prime)|^2(\psi^\ast(\vec{r}^\prime)R_2
(\vec{r}
,\vec{r}^\prime)
 + \psi(\vec{r}^\prime)R_1^\ast(\vec{r}, \vec{r}^\prime)) d\vec{r}^\prime = -\psi(\vec{r}), \\
&~& \int~g_0N_0|\psi(\vec{r}^\prime)|^2(\psi^\ast(\vec{r}^\prime)R_1(\vec{r}, \vec{r}^\prime) + \psi(\vec{r}^\prime)R_2^\ast(\vec{r}, \vec{r}^\prime))d\vec{r}^\prime = 0 .
\end{array}
\eqno{(41)}
$$

\noindent
Eqs. (41) do not have solutions if $\psi(\vec{r}) \not= 0$.  As noted in
[36], when constructing variational principles for certain quantities (as,
for example, for $\psi(\vec{r})$), it is expected that some 
specification of the phase of the wave function must be imposed; otherwise,
various ambiguities and contradictions can occur (see Eqs. (40)).  There
are various ways of specifiying the phase of $\psi$.  The simplest and
useful procedure is to fix the phase of $\psi$ relative to some arbitrary
known function $\chi(r)$ through a restriction of the sort that $<\psi|\chi>$
is either purely real or purely imaginary [36].
\vspace{8pt}

Therefore we write
$$\begin{array}{rcl}
[\psi](\vec{r}) &=& \psi_t(\vec{r}) + \gamma_t(\vec{r})(<\psi_t|\chi> -
<\chi|\psi_t>) + \\
&+& \int~R_2^{t \ast}(\vec{r}, \vec{r}^\prime)[T + V_{trap} + g_0N_0|\psi_t|^2 
- \mu_t]\psi_t(\vec{r}^\prime)d\vec{r}^\prime + \\
&+& \int[(T + V_{trap} + g_0N_0|\psi_t|^2 - \mu_t)\psi_t(\vec{r}^\prime)]^\ast R_1^t(\vec{r},
\vec{r}^\prime)d\vec{r}^\prime, 
\end{array}
\eqno{(42)}
$$

\noindent
where
$$\begin{array}{rcl}
(T + V_{trap} &+& 2g_0N_0|\psi_t|^2 -\mu_t)R_2^{t \ast}(\vec{r}, \vec{r}^\prime) + g_0N_0\psi_t^{\ast 2} R_1^t(\vec{r}, \vec{r}^\prime) + \\
&+& \delta(\vec{r} - \vec{r}) - \gamma_t(\vec{r})\chi^\ast(\vec{r}^\prime) = 0, 
\end{array}
\eqno{(43)}
$$

\noindent
$$\begin{array}{rcl}
(T &+& V_{trap} + 2g_0N_0|\psi_t|^2 - \mu_t)R_1^t(\vec{r}, \vec{r}^\prime) +
g_0N_0 \psi_t^2 R_2^{t \ast}, (\vec{r}, \vec{r}^\prime) + \\
&+& \gamma_t(\vec{r}) \chi(\vec{r}^\prime) = 0 , 
\end{array}
\eqno{(44)}
$$

\noindent
$$\begin{array}{rcl}
~& \gamma_t(\vec{r}) =0.5 \psi_t(\vec{r})/<\psi_t|\chi> ,
\end{array}
\eqno{(45)}
$$

\noindent
and
$$ <\psi_t|\chi> = <\chi|\psi_t> .
\eqno{(46)}
$$

Varying Eq. (42) we obtain 
$$ \delta[\psi](\vec{r}) = O(\delta \psi^2) ,
\eqno{(47)}
$$

\noindent
where we have assumed Eq. (31) and
$$ \delta R_i = O(\delta \psi), i = 1, 2
\eqno{(48)}
$$

\noindent
Using Eqs. (43)-(46) we can rewrite Eq. (42) as Eq. (37).

\noindent
{\bf 6. Variational-iteration method}

\vspace{5pt}

To solve nonlinear problems, one frequently uses the method of iteration, several
iteration schemes are possible [15, 37].
\vspace{5pt}

Now we formulate iteration method which directly follows from variational methods
given by Eqs. (37)and (38),
$$\begin{array}{rcl}
 \psi^{(N)}(\vec{r}) &=& \int~d \vec{r}^\prime Q_1^{(N-1)}(\vec{r}, \vec{r}^\prime)
\psi^{(N-1)}(\vec{r}^\prime)d \vec{r}^\prime \\

 &+&  \int d \vec{r}^\prime Q_2^{(N-1)}(\vec{r}, \vec{r})(\psi^{(N-1)}(\vec{r}^\prime))^\ast d \vec{r}^\prime , 
\end{array}
\eqno{(49)}
$$

\noindent
where, for example, for ground state $(\psi^{(N)} = \psi^{(N)\ast})$, 
$$ \begin{array}{rcl}
    Q_1^{(N-1)} (\vec{r}, \vec{r}^\prime) &=& Q_2^{(N-1)} (\vec{r}, \vec{r}^\prime) \\

     &=& g_0N_0 \psi^{(N-1)}(\vec{r}^\prime) L_R^{(N-1)} (\vec{r}, \vec{r}^\prime).
\end{array}
\eqno{(50)}
$$

\noindent
$L_R^{(N-1)}(r, r^\prime)$ is a solution of the linear equation 
$$ \begin{array}{rcl}
 [T + V_{trap} &+& 2g_0N_0|\psi^{(N-1)}|^2 - \mu^{(N-1)}]L_R^{(N-1)} + g_0N_0
(\psi^{(N-1)})^2 L_R^{(N-1)}  \\

 &=& \delta (\vec{r} - \vec{r}^\prime) ,
\end{array}
\eqno{(51)}
$$

\noindent
where 
$$ \begin{array}{rcl}
 \mu^{(N-1)} &=& \mu_t^{(N-1)} + \lambda^{(N-1)} \langle L^{(N-1)}|T + V_{trap} + g_0N_0|\psi^{(N-1)}|^2
 - \mu_t^{(N-1)}|\psi^{(N-1)} \rangle , \\
 \lambda^{(N-1)} &=& 1/ \langle L^{(N-1)}|\psi^{(N-1)} \rangle,
\end{array}
\eqno{(52)}
$$

$$ \mu_t^{(N-1)} = \langle \psi^{(N-1)}|T + V_{trap} + V_H^{(N-1)}|\psi^{(N-1)} \rangle,
\eqno{(53)}
$$
\noindent
with
$$ V_H^{(N-1)} = g_0N_0|\psi^{(N-1)}(\vec{r})|^2 ,
\eqno{(54)}
$$

\noindent
and $L^{(N-1)}$ is a quadratically integrable solution of the linear nonhomogeneous equation
$$ (T +  V_{trap} + 3 V_H^{(N-1)} - \mu_t^{(N-1)}) L^{(N-1)} = -\psi^{(N-1)} .
\eqno{(55)}
$$

Eqs. (50-55) can be easily generalized to include vortex states. We
note that
one of the primary motivations of the GPG theory was the study of vortex states
in weakly interacting bosons [30, 31].  We expect a super-fast convergence of the
iteration process describing by Eqs. (49-55).  It follows from the fact that, if $(\psi - \psi^{(N-1)})
= O(\delta \psi)$, then $(\psi - \psi^{(N)})) = O(\delta \psi^2)$.
\vspace{5pt}

To simplify the iteration process given by Eqs. (49-55) let us introduce 
$$ \begin{array}{rcl}
 \psi^{(N)} (\vec{r}) &=& \int d \vec{r}^\prime Q_1^{(0)} (\vec{r}, \vec{r}^\prime)
 \psi^{(N-1)}(\vec{r}^\prime) d\vec{r}^\prime \\

 &+& \int d \vec{r}^\prime Q_2^{(0)} (\vec{r}, \vec{r}^\prime)(\psi^{(N-1)}(r^\prime))^\ast d \vec{r}^\prime  .
\end{array}
\eqno{(56)}
$$

\noindent
This scheme is much simpler than Eqs. (49 - 55) since we need to solve a set of
linear equations (51) and (55) only once (for N = 1).   If function $\psi^{(0)}$ is
chosen sufficiently close to $\psi$, the $Q_i^{(N-1)} (r, r^\prime)$ and $Q_i^{(0)}
(\vec{r}, \vec{r}^\prime)$ will differ only by small amount.  This provides the
basis for the iteration process given by Eq. (56) [36]. To solve a set of linear
equations, Eqs. (51) and (55), we can expand solutions in a basis of anisotropic
trap eigenfunctions [15].
$$ \begin{array}{rcl}
 \psi^{(N-1)} (\vec{r}) &=& \sum_{i=1}^{N_{basis}}~\alpha_i^{(N-1)} \phi_{x_i}(x) \phi_{y_i}(y)
 \phi_{z_i}(z) , \\

  L^{(N-1)}(\vec{r}) &=& \sum_{i=1}^{N_{basis}}~\beta_i^{(N-1)} \phi_{x_i}(x) \phi_{y_i}(y)
\phi_{z_i} (z), \\

  L_R^{(N-1)}(\vec{r}, \vec{r}^\prime) &=& \sum_{i,j, = 1}^{N_{basis}} \gamma_{ij}^{(N-1)} 
 \phi_{x_i} (x) \phi_{y_i}(y) \phi_{z_i} (z) \phi_{x_j} (x^\prime) \phi_{y_j}(y^\prime)
 \phi_{z_j} (z^\prime),
\end{array}
\eqno{(57)}
$$

\noindent
where $\phi_{n_i}(y)$ is a one-dimensional harmonic-oscillator eigenfunction.  The
coefficients, $\alpha_i^{(N-1)}, \beta_i^{(N-1)}$, and $\gamma_{ij}^{(N-1)}$, can be
expressed in terms of 
$$ J_{n_i n_j n_k n_l} = \int_{-\infty}^{+\infty}~dy \phi_{n_i} (y) \phi_{n_j} (y)
 \phi_{n_k} (y) \phi_{n_l}(y),
\eqno{(58)}
$$

\noindent
which can be evaluated analytically [38].
\vspace{8pt}

{\bf 7.  Summary and Conclusions}
\vspace{5pt}

Based on the GRS approach [36], we have
derived variational principles for the nonlinear Schr\"odinger
equation in a systematic way.  In particular, we have obtained
variational principle for eigenvalues (chemical potential) and
wave function of the GPG equation.
To the best of our knowledge, these variational
principles for the GPG equation are new.  Using these variational
principles, we have formulated a variational iteration method,
which is expected to have a very fast convergence.
\vspace{8pt}

We are planning to generalize our results to the case of nonstationary
GPG equation and to carry out relevant numerical calculations based
on our method.
\pagebreak

\begin{center}
{\bf Appendix}
\end{center}
Our variational iteration method developed in this paper is
general and can be applied to both nonlinear differential
equation and nonlinear integral equation.
In this Appendix, we describe one example of the application
of our variational iteration method to non-linear integral
equation
$$  y(x) = \int_a^b K(x,s,y(s))ds,~~~(x,s)\epsilon [a,b],
   \eqno{(A.1)}
$$
where $K(x,s,y)$ is a continuous function, with all its
variables continuous, and has continuous low order
derivatives.  We introduce a functional $[y](x)$ whose
stationary value is $y(x)$
$$  [y](x) =\int_a^b K(x,t,y_t(t))dt +
    \int_a^b L(x,x^\prime)[y_t(x^\prime) - \int_a^b K
        (x^\prime,t,y_t(t))dt]dx^\prime ,
 \eqno{(A.2)}
 $$
where $L(x,x^\prime)$ is solution of a linear integral equation
$$  L(x,t) - \int_a^b ds^\prime L(x,s^\prime) \frac{\partial K}
                                            {\partial y_t}
        (s^\prime,t,y_t(t)) = - \frac{\partial K(x,t,y_t(t))}
                                      {\partial y_t}  .
\eqno{(A.3)}
$$
Iteration procedure which follows directly from the
variational functional (A.2) is
$$ \begin{array}{rcl}
 y^{(N)}(x) &=& \int_a^b K(x,t,y^{(N-1)}(t)) dt \\
&+& \int_a^b L^{(N-1)}(x,s) [y^{(N-1)}(s) \\
 &-& \int_a^b K(s,t,y^{(N-1)}(t))dt]ds,
\end{array}
\eqno{(A.4)}
$$

$$  L^{(N-1)}(s,t) - \int_a^b dx^\prime L^{(n-1)}
    (s,x^\prime) \frac{\partial K(x^\prime,t,y^{(N-1)})}
                      {\partial y^{(N-1)}}
      =- \frac{\partial K(s,t,y^{(N-1)}(t))}
                     {\partial y^{(N-1)}}               .
\eqno{(A.5)}
$$

\noindent
The above iteraction procedure is equivalent to the
Kantorivich-Newton process (KN) [37] which has developed for
nonlinear integral equations:

$$  y^{(N)}(x) = y^{(N-1)}(x) + \phi_{N-1}(x),
   \eqno{(A.6)}
  $$

 $$  \phi_{N-1}(x) = \epsilon_{N-1}(x) + \int_a^b
     \frac{\partial K(x,s,y^{(N-1)}(s))}
          {\partial y^{(N-1)}}
      \phi_{N-1} (s)ds,
   \eqno{(A.7)}
 $$

\noindent
where
$$  \epsilon_{N-1}(x) = \int_a^b K(x,s,y^{(N-1)}(s))ds -
       y^{(N-1)}(x).
   \eqno{(A.8)}
$$

 \noindent
 \vspace{8pt}

To show the equivalence between the KN method and our method for
solving the nonlinear integral equation (A.1), we rewrite
solution of Eq. (A.7) as
$$ \phi_{N-1}(x) = \int_a^b~\Gamma^{(N-1)}(x,s)\epsilon_{N-1}(s)ds,
\eqno{(A.9)}
$$

\noindent
where
$$ \Gamma^{(N-1)}(x,t) = \delta(x-t) + \int_a^b~\frac{\partial
K(x,s, y{(N-1)}(s)}{\partial y^{(N-1)}}~\Gamma^{(N-1)}(s,t)ds .
\eqno{(A.10)}
$$

\noindent
Substitution of Eq. (A.9) into Eq. (A.6) gives
$$\begin{array}{rcl}
 y^{(N)}(x) &=& \int_a^b K(x,t,y^{(N-1)}(t))dt\\
&+& \int_a^b[-ds^\prime
\frac{\partial K(x,s^\prime,y^{(N-1)}(s^\prime))}{\partial
y^{(N-1)}}\Gamma^{(N-1)}(s^\prime,s)](y^{(N-1)}(s) - \\
&-& \int_a^b~K(s,t,y^{(N-1)}(t)dt)ds
\end{array}
\eqno{(A.11)}
$$

\noindent
Using Eq. (A.10), we obtain
$$ - \int_a^b ds^\prime \frac{\partial K(x,s^\prime,
y^{(N-1)}(s^\prime}{\partial y^{(N-1)}}\Gamma^{(N-1)} 
(s^\prime, s) = L^{(N-1)}(x,s)
\eqno{(A.12)}
$$

\noindent
Therefore, the KN method developed for the nonlinear integral
equation (A.1) is a special case of the GRS approach.
\vspace{8pt}

We shall illustrate a numerical convergence of the iteration
procedures (A.4)-(A.5) for an example [37] of the
     following nonlinear integral equation,
$$  y(x) = \frac{1}{2} \int_0^1 [y(t)]^2 x tdt + 1 .
   \eqno{(A.13)}
$$
The exact solution of this equation has the form
$$  y(x) = 1 + ax,
   \eqno{(A.14)}
 $$
where a is given as solutions of a quadratic equation.
There are two solutions of Eq. (A.9) with
$a \approx 0.40589$ and $a \approx 4.92744$.

For case of $y^{(N)} = 1 + a^{(N)} x$, the iteration procedure
(A4-A5) reduces to
$$ \begin{array}{rcl}
 a^{(N)} &=& [\frac{1}{4} + \frac{1}{3} a^{(N-1)}
      + \frac{1}{8} (a^{(N-1)})^2]
     [1 - \frac{12}{3a^{(N-1)}-8} (1/3 + \frac{a^{(N-1)}}{4})]
\\
    &+& \frac{12}{3a^{(N-1)}-8} a^{(N-1)}
  (\frac{1}{3} + \frac{a^{(N-1)}}{4}).
\end{array}
\eqno{(A.15)}
$$

\noindent
Starting with $a^{(0)} = 0$, let us calculate the differences
$\Delta^{(N)}$ defined as

$$  \Delta^{(N)} = \frac{a_{exact}-a^{(N)}}
                        {a_{exact}},
   \eqno{(A.16)}
$$
where
 $$ a_{exact} = \frac{8}{3} - ((\frac{8}{3})^2-2)^{1/2}.
$$
The results are summarized in the Table I.
\vspace{8pt}

 From Table I, we can see a super-fast convergence
of the procedure (A4-A5) for the nonlinear integral equation
(A.13).
\vspace{8pt}

Starting with values of $a^{(0)}$ in range of $0 \leq a^{(0)}
\leq 10$, we have  calculated the corresponding values of
$a^{(N)}$.  The results are summarized in the Table II. We note
that in general the convergence of iterational procedure depends
on initial step. If $a^{(0)}$ is chosen sufficiently close to the
exact value we have a super-fast convergence.
\pagebreak

\begin{center}
{\bf Table I}
\end{center}
\vspace{12pt}

\begin{center}
\begin{tabular}{|c|ccccc|}
\hline
N & 0 & 1 & 2 & 3 & 4\\
\hline
$\Delta^{(N)}$ & 1 & $7\times 10^{-2}$ & $5\times 10^{-4}$ &
       $2 \times 10^{-8}$ & $5 \times 10^{-17}$ \\
\hline
\end{tabular}
\end{center}
\vspace{20pt}

\begin{center}
{\bf Table II}
\end{center}
\vspace{12pt}

\begin{center}
\begin{tabular}{|c|c|c|c|c|c|}
\hline
$a^{(0)}$ & $a^{(1)}$ & $a^{(2)}$ & $a^{(3)}$ & $a^{(4)}$ &
$a^{(5)}$ \\
\hline
0 & 0.37500 & 0.40568 & 0.40589 & 0.40589 & 0.40589 \\
\hline
1 & 0.30000 & 0.40352 & 0.40588 & 0.40589 & 0.40589 \\
\hline
2 & -l.50000 & -0.03000 & 0.37066 & 0.40562 & 0.40589 \\
\hline
3 & 10.50000 & 6.90957 & 5.39040 & 4.96679 & 4.92778 \\
\hline
4 & 5.25000 & 4.94758 & 4.92753 & 4.92744 & 4.92744 \\
\hline
5 & 4.92857 & 4.92744 & 4.92744 & 4.92744 & 4.92744 \\
\hline
6 & 5.10000 & 4.93356 & 4.92745 & 4.92744 & 4.92744 \\
\hline
7 & 5.42308 & 4.97200 & 4.92787 & 4.92744 & 4.92744 \\
\hline
8 & 5.81250 & 5.05190 & 4.93069 & 4.92745 & 4.92744 \\
\hline
9 & 6.23684 & 5.16756 & 4.93897 & 4.92747 & 4.92744 \\
\hline
10 & 6.68182 & 5.31072 & 4.95522 & 4.92761 & 4.92744 \\
\hline
\end{tabular}
\end{center}
\pagebreak

\begin{center}
{\bf References}
\end{center}
\vspace{8pt}

\noindent
[1]  S. N. Bose, Z. Phys. {\bf 26}, 178 (1924); A. Einstein, Sitz.

Preuss Akad. Wiss. {\bf 1924}, 3 (1924).
\vspace{5pt}

\noindent
[2]  A. Griffin, D. Snoke, and S. Stringari, {\it Bose-Einstein Condensation}

(Cambridge, New York, 1995).
\vspace{5pt}

\noindent
[3]  M. H. Anderson, J. R. Ensher, M. R. Matthews, C. E. Wieman,

and E. A. Cornell, Science {\bf 269}, 198 (1995).
\vspace{5pt}

\noindent
[4]  C. C. Bradley, C. A. Sackett, J. J. Tollett, and R. G. Hulet,

Phys. Rev. Lett. {\bf 75}, 1687 (1995).  In this work, the

condensate could not be directly observed. See C. C. Bradley, C.

A. Sackett, and R. G. Hulet, Phys. Rev. Lett. {\bf 78}, 985

(1997).

\vspace{5pt}

\noindent
[5]  K. B. Davis, M. -O. Mewes, M. R. Andrew, N. J. van Druten, D. S.

Durfee, D. M. Kurn, and W. Ketterle, Phys. Rev. Lett. {\bf 75},

3969 (1995).
\vspace{5pt}

\noindent
[6] M. -O. Mewes, M. R. Andrews, N. J. van Druten, D. M. Kurn, 

D. S. Durfee, and W. Ketterle, Phys. Rev. Lett {\bf 77}, 416

(1996).
\vspace{5pt}

\noindent
[7] D. S. Jin, J. R. Ensher, M. R. Matthews, C. E. Wieman, and E. A. 

Cornell, Phys. Rev. Lett. {\bf 77}, 420 (1996).
\vspace{5pt}

\noindent
[8] M. -O. Mewes, M. R. Andrews, N. J. van Druten, D. M. Kurn, D. S.

Durfee, C. G. Townsend, and W. Ketterle, Phys. Rev. Lett. {\bf 77},

988 (1996).
\vspace{5pt}

\noindent
[9] J. R. Ensher, D. S. Jin, M. R. Matthews, C. E. Wieman, and E. A. 

Cornell, Phys. Rev. Lett. {\bf 77}, 4984 (1996).
\vspace{5pt}

\noindent
[10] C. J. Myatt, E. A. Burt, R. W. Ghrist, E. A. Cornell, and C. E. Wieman, 

Phys. Rev. Lett. {\bf 78}, 586 (1997).
\vspace{5pt}

\noindent
[11] D. S. Jin, M. R. Matthews, J. R. Ensher, C. E. Wieman, E. A. Cornell, 

Phys. Rev. Lett {\bf 78}, 764 (1997).
\vspace{5pt}

\noindent
[12]  Mark Edwards and K. Burnett, Phys. Rev. {\bf 51}, 1382 (1995).
\vspace{5pt}

\noindent
[13]  P. A. Ruprecht, M. J. Holland, K. Burnett, and Mark Edwards, Phys.

Rev. {\bf 51}, 4704 (1995).
\vspace{5pt}

\noindent
[14]  G. Baym and C. J. Pethick, Phys. Rev. Lett. {\bf 76}, 6 (1966). 
\vspace{5pt}

\noindent
[15]  Mark Edwards, D. J. Dodd, C. W. Clark, P. A. Ruprecht, and K.

Burnett, Phys. Rev. A {\bf 53}, R1950 (1966). 
\vspace{5pt}

\noindent
[16] F. Dalfovo and S. Stringari, Phys. Rev. A {\bf 53}, 2477 (1996).
\vspace{5pt}

\noindent
[17] A. Griffin, Phys. Rev. B {\bf 53}, 9341 (1996). 
\vspace{5pt}

\noindent
[18] A. L. Fetter, Phys. Rev. A {\bf 53}, 4245 (1996).
\vspace{5pt}

\noindent
[19]  R. J. Dodd, Mark Edwards, C. J. Williams, C. W. Clark, M. J. Holland, 

P. A. Ruprecht, and K. Burnett, Phys. Rev. A {\bf 54}, 661 (1996).
\vspace{5pt} 

\noindent
[20]  F. Dalfovo, C. Minniti, S. Stringari, L. Pitaevskii, Phys. Lett.

A227, 259 (1977). 
\vspace{5pt}

\noindent
[21]  Y. Kagan, G. V. Shlyapnikov, and J. T. M. Walraven, Phys. Rev. Lett. 

{\bf 76}, 2670 (1996).
\vspace{5pt}

\noindent
[22]  Mark Edwards, P. A. Ruprecht, K. Burnett, R. J. Dodd, and C. W.

Clark, Phys. Rev. Lett. {\bf 77}, 1671 (1996).
\vspace{5pt} 

\noindent
[23]  E. M. Wright, D. F. Walls, and J. C. Garrison, Phys. Rev. Lett. {\bf 77},

2158 (1966). 
\vspace{5pt}

\noindent
[24]  S. Stringari, Phys. Rev. Lett. {\bf 77}, 2360 (1996). 
\vspace{5pt}

\noindent
[25] V. M. P\'erez-Garc\'ia, H. Michinel, J. I. Cirec, M. Lewenstein, and P. 

Zoller, Phys. Rev. Lett. {\bf 77}, 5320 (1996).
\vspace{5pt}

\noindent 
[26]  S. Giorgini, L. P. Pitaevskii, and S. Stringari, Phys. Rev. A {\bf 54}, R4633
 
(1996).

\noindent
[27] B. D. Esry, Phys. Rev. A {\bf 55}, 1147 (1997).
\vspace{5pt}

\noindent
[28] A. L. Fetter, J. Low Temp. Phys. {\bf 106}, 643 (1997).
\vspace{5pt}

\noindent
[29] V. L. Ginsburg and L. P. Pitaevskii, Sov. Phys. JETP 7, 858 (1958).
\vspace{5pt}

\noindent
[30] L. P. Pitaevskii, Sov. Phys. JETP {\bf 13}, 451 (1961).
\vspace{5pt} 

\noindent
[31] E. P. Gross, Nuovo Cimento {\bf 20}, 454 (1961); J. Math. Phys. {\bf 4},

195 (1963). 
\vspace{5pt}

\noindent
[32] A. L. Fetter and T. D. Walecka, {\it Quantum Theory of many-particle

system} (McGraw-Hill, 1971).
\vspace{5pt}

\noindent
[33] V. L. Ginzburg, Physics-Uspekhi 40, 407 (1997).
\vspace{5pt}

\noindent
[34]  L. Vazques, L. Streit and V. M. P\'erez-Garc\'ia, {\it Nonlinear Klein-Gordon

and Schr\"odinger Systems:  Theory and Applications} (World Scientific, 

(1996); M. J. Allowitz and P. A. Clarkson, {\it Solitons}, {\it Nonlinear 

Evaluations and Inverse Scattering} (Cambridge Univ. Press, New

York, 1991).
\vspace{5pt}

\noindent
[35] A. L. Zubarev, Sov. J. Part. Nucl. {\bf 4}, 188 (1978).
\vspace{5pt}

\noindent
[36] E. Gerjuoy, A. R. P. Rau, and L. Spruch, Rev. Mod. Phys. {\bf 725}

(1983). 
\vspace{5pt}

\noindent
[37] L. V. Kantorivich and G. P. Akilov, {\it  Functional analysis

in Normed Spaces} (MacMillan, New York, 1964), Chapter 18.
\vspace{5pt}

\noindent
[38] I. W. Burbridge, J. London. Math. Soc. {\bf 23}, 135 (1948).

\end{document}